\documentclass[12pt,preprint]{aastex}

\def\pd #1/#2{\frac{\partial #1}{\partial #2}}
\def\v(#1){{\bf #1}}
\def\div {\nabla\cdot}
\def\del {\nabla}
\def\half {{{1}\over{2}}}
\def\ie {i.e.}
\def\implies{\Rightarrow}

\shortauthors{Delamarter, Frank, and Hartmann}
\shorttitle{Interaction of Infall and Winds in YSOs}

\usepackage{afterpage,wrapfig,epsfig,graphicx}
\begin{document}
%\psdraft
\singlespace
\title{Interaction of Infall and Winds\\
    in Young Stellar Objects}

\author{G. Delamarter$^1$, A. Frank$^1$ \& L. Hartmann$^2$}
\affil{$^1$ Department of Physics and Astronomy, University of Rochester,
    Rochester, NY 14627}
\affil{$^2$ Harvard-Smithsonian Center for Astrophysics, Cambridge, MA 02138}

\begin{abstract}

  The interaction of a stellar or disk wind with a collapsing
  environment holds promise for explaining a variety of outflow
  phenomena observed around young stars.  In this paper we present the
  first simulations of these interactions.  The focus here is on
  exploring how ram pressure balance between wind and ambient gas and
  post-shock cooling affects the shape of the resulting outflows. In
  our models we explore the role of ram pressure and cooling by holding
  the wind speed constant and adjusting the ratio of the inflow mass
  flux to the wind mass flux ($\dot{M}_a/\dot{M}_w$)  Assuming
  non-spherical cloud collapse, we find that relatively strong winds
  can carve out wide, conical outflow cavities and that relatively weak
  winds can be strongly collimated into jet-like structures.  If the
  winds become weak enough, they can be cut off entirely by the
  infalling environment.  We identify discrepancies between results
  from standard snowplow models and those presented here that have
  important implications for molecular outflows.  We also present mass
  vs.  velocity curves for comparison with observations.

\end{abstract}


\section{Introduction}

Molecular outflows are commonly observed in association with young
stars.  Although the precise mechanism generating the outflows is
poorly understood, it is generally believed that the molecular mass is
driven and excited into emission by a wind emanating from the inner
circumstellar disk close to the star, and driven by accretion energy
\citep{CalvetGullbring1998}.  Recent observations have explored
how these outflows form close to the driving source
\citep{ChandlerEtAl96} and how they might evolve in time
\citep{VelusamyLanger98}.  These and other investigations indicate
that molecular outflows are not clearly explained in terms of any
single current model.

Molecular outflow models are principally divided into two groups;
those in which outflows are driven by highly-collimated winds or
``jets'', and those in which outflows are driven by wide-angle winds.
\citet{MassonChernin92,MassonChernin93} and
\citet{CherninMasson95} compared jet driven and wind driven
models of molecular outflows. They conclude that jet-driven models are
a better fit than wide-angle wind models because observations show
very little outflow mass moving at the highest velocities.  A jet with
moderate post-shock cooling would transfer its momentum to the ambient
molecular material largely through the small cross-section of its
head, while a wide wind has a much larger driving area.  On the other
hand, jet-driven models have so far had some difficulty producing the
cross-sectional widths of many observed outflows.  To reproduce these
morphologies, investigators have invoked mechanisms such as turbulent
entrainment \citep{Stahler94,RagaCabrit93}, and precession
\citep{CliffeFrankEtAl95,SuttnerSmithEtAl1997}.  These strategies
imply filled outflow cavities and shock structures that aren't
generally observed \citep{ChandlerEtAl96}.  Wide angle wind models
can produce wind-blown bubbles with wide bow shocks and empty
cavities.  It is, however, difficult for the simplest versions of
these models to reproduce the observed momentum distribution.
\citet{LiShu96} have suggested that a wide angle wind whose
properties vary as a function of polar angle may help solve the
momentum problem.  Such a wind requires a structure that has higher
density along an axis: more like a jet.  Thus at present, models and
observations leave it unclear as to what is really the driving
mechanism of molecular outflows.  With these issues in mind, the
present paper examines a wide angle wind model through hydrodynamic
simulations.
  
The early model used by Chernin and Masson to make their case against
wide angle winds is that presented by \citet{ShuRudenLadaLizano91}.  
It applies a ``snowplow'' scenario
where a central wind sweeps up ambient material and compresses it into
a thin shell.  The calculation is performed under the critical
assumptions of isothermal shock dynamics and strong mixing between the
post-shock wind and ambient gas.  The model successfully produces
collimated bipolar wind-blown bubbles when relatively denser
``equatorial'' regions prevent the shell from expanding at low
latitudes as quickly as it does along the pole.
  
New features have been added to the basic wide angle-wind blown bubble
model.  Close to the young star, effects such as gravity from the
central star and the density distribution of the inner envelope become
crucial in influencing the shape and evolution of the outflow.  This
has been investigated in detail by \citet{WilkinStahler98}, who
model a quasi-static balance between a steady central wind, gravity,
and a time-dependent angularly varying infall.  In essence the bubble
is considered as a series of accretion shocks.  Using the momentum
balance across the fixed shell face in both normal and transverse
directions, the authors solve for the density and flow pattern within
the shell. Because of the quasi-static assumption, the time-dependence
of the bubble shape and size is directly linked to the way in which
the infall changes in time.  The stability of such a situation remains
to be investigated.  While they found that significant collimation can
occur as the environmental density distribution becomes more
flattened, the timescale for this is $\sim10^5$ yr.  They point out
that this is much longer than what is observed.  This result reveals
the importance of performing dynamic, time-dependent calculations.

While considerable progress has been made in the wide angle wind
scenario for molecular outflows there remain many aspects which have
not yet been explored. In particular the full time-dependent
multi-dimensional nature of the flows has not been examined.  Solving
cylindrically symmetric non-linear, time-dependent fluid equations by
numerical simulation we obtain significantly different behavior from
that obtained in previous wide angle wind blown bubble models.  The
shape of the envelope, namely the gradient of density with polar
angle, not only affects the shape of the outer shock and resulting
molecular shell, but also the shape of the inner (wind) shock.  This
can have important dynamical consequences as has been shown in
\citet{FrankMellema96} and \citet{MellemaFrank97}.

It seems difficult to avoid the the conclusion that some intrinsic
collimation of the initial mass ejection is required, given the
evidence for highly-collimated jets over long scales
\citep{BallyEtAl1997,Bachiller96}.  This is particularly true for
emission-line jets of optically-visible T Tauri stars (e.g.,
\citealt{StapelfeldtEtAl1997}; see also the review by
\citealt{ReipurthHeathcote1997}), where there is little evidence
for the presence of enough external (dusty) medium for initial
hydrodynamic collimation of ionized jets to be effective.  On the
other hand, it should also be recognized that even models of mass
ejection with collimated flows often have some wide-angle component.
This is true not only of the magnetic ``X-wind'' model of Shu and
collaborators \citep{ShuEtAl94,NajitaShu94,ShangEtAl1998}, in
which a dense axial flow is surrounded by a lower-density dispersing
flow; it is also true of the original self-similar MHD disk wind
\citep{BlandfordPayne82,PudritzNorman83,Konigl89}.  For this
reason it is worth exploring the interaction of wide-angle flows with
the ambient medium.

In this paper we consider the time-dependent interaction of an
infalling envelope with an expanding wind.  Our focus is
spherically-symmetric winds as an example of a maximally wide angle
wind, though we do calculate two cases
using aspherical winds for
comparison.  Collimation of the resulting flow will be
enhanced if the infalling envelope is not spherically-symmetric.  We
take one of the simplest models for such non-spherical collapse, the
model of \citet{HartmannCalvetBoss96}, from an initial
self-gravitating sheet with no magnetic field.

In \S~\ref{sec:Computation} we present the physics and the
computational scheme used for the simulations.  The simulation results
are examined in \S~\ref{sec:Results}.  In \S~\ref{sec:Discussion} we
compare our simulations to the snowplow model of outflows, discuss the
possible role and importance of turbulence in these models, and make
comparisons of the simulations to observations.  We conclude in
\S~\ref{sec:Conclusion}.

\section{Computation}
\label{sec:Computation}

\subsection{Methods}

The simulation code solves the Euler ideal fluid equations with source
(sink) terms for central point-source gravity and radiative
cooling.  The equations, written in conservative form, are:

\begin{eqnarray}
\pd {\rho}/{t}+\div\rho\v(u)=0, \\
\pd {\rho\v(u)}/{t}+\div\rho\v(u)\v(u)=\rho\del\Phi, \\
\pd {\epsilon}/{t}+\div\v(u)(\epsilon+p)=-\Lambda(\rho,T)-\rho\v(u)\cdot\del\Phi.
\end{eqnarray}

\noindent where $\rho$ is the fluid mass density at a point, and $\v(u)$ is the
velocity. The pressure $p$ is related to the energy density $\epsilon$
and its kinetic ($\epsilon_{kn}$) and thermal components
($\epsilon_{th}$) through the relations
\begin{eqnarray}
\epsilon=\epsilon_{kn}+\epsilon_{th},\\
\epsilon_{kn}=\half\rho u^2, \\
\epsilon_{th}={{p}\over{(\gamma-1)}}.
\end{eqnarray}

\noindent The adiabatic index $\gamma=5/3$ is that for a monatomic
gas.  The temperature is determined through the ideal gas equation of
state, with the particle mass set to that of atomic hydrogen:  
\begin{equation} 
p={{\rho k T}\over{m_H}}.  
\end{equation} 
\noindent
The cooling source term $\Lambda(\rho,T)$ includes a Dalgarno-McCray
\citeyear{DM72} cooling curve which is applied above 6000 K, and a
Lepp and Shull \citeyear{LeppShull83} cooling curve applied below 6000
K to simulate the effects of both high and low temperature cooling.

\noindent Gravity in these models is purely due to a source at the
origin. Self-gravity of the fluid is not included.  The
gravity source term is written in terms of the gradient of the
potential $\del\Phi$.  This force is specified via the constant central
source mass $M_*$ as
\begin{equation} 
\del\Phi=G {{M_*}\over{r^2}}\hat r,
\end{equation}
\noindent The radius $\v(r)$ is the vector from the origin to the fluid point.

The equations are solved on an Eulerian grid using an operator split
numerical method, where several operators are applied per timestep,
each one simulating a different aspect of the physics.  We also employ
fluid tracking to determine what part of the initial grid a fluid
parcel came from.  The grid covers a quarter meridional plane of a
cylinder and has assumed axial symmetry plus reflective symmetry
across an equatorial plane.  The basic hydrodynamics equations without
source terms are applied via the numerical Total Variation Diminishing
(TVD) method of Harten \citeyear{Harten83} as implemented in
dimensionally split form by \citet{RyuEtAl95}.  The source term
for radiative cooling is applied explicitly to first order via an
exponential as described in \citet{MellemaFrank97} where:
\begin{equation}
\epsilon_{th}^{n+1}=\epsilon_{th}^{n}\exp\left(-{{\Lambda^n(\rho^n,T^n)}\over{\epsilon_{th}^n}}\Delta
t\right).
\end{equation}

\noindent The superscripts represent a value at the $n$th and
$(n+1)$th timestep, and $\Delta t$ represents the amount of time
between timestep $n$ and $n+1$.  Gravity is applied through a first
order explicit Euler operator which updates the velocities and the
kinetic energy density.  The fluid is evolved by rotational velocity
through another first order explicit Euler operator.

The timesteps are adjusted to satisfy the Courant condition on the
sourceless hydrodynamics, and simultaneously to be less than a
reasonable multiple of the cooling time:
$3{{\Lambda(\rho,T)}/{\epsilon_{th}}}$. Because of the high densities
in the ambient medium ($n > 10^8 ~cm^{-3}$) the cooling time step can
become relatively small and the number of times steps required to
complete the simulation relatively large. This produces additional
diffusion in the solutions causing our flows to appear rather smooth.
By artificially reducing the cooling in some test simulations, we were
able to obtain sharper features. The gross morphology and evolution of
the flow however remained qualitatively the same as those with the
full cooling shown in \S~\ref{sec:Results}.

We applied several tests to the code.  We compared momentum conserving
wind blown bubbles with the solutions obtained by
\citet{KooMcKee92B} to test the basic hydrodynamics with strong
cooling.  We performed radiative shock tube tests along the axis using
a simplified power law cooling function and compared these against
semi-analytic shock solutions to test the cooling operator.  To test
gravity we reproduced the Bondi accretion solution in the code.  To
test gravity and rotational velocity, we maintained a Keplerian ring
in an orbit around a central mass.  In general, the code was able to
recover the analytical solutions to better than 10\%.

\subsection{Model}

The scenario we applied involved driving a spherically symmetric wind
into an infalling non-spherical envelope.  While a spherical wind may
not be operating in many young stellar object systems, we seek in this
investigation to explore the shaping of the outflow through
wind/environment interactions and make contact with previous analytic
studies.  We hope to relax the assumption of a spherical wind in a
later paper.  The parameters used are given in Table
\ref{table:simparam} and are based on typical values for low mass
young stellar objects.  A radially directed steady wind of velocity
$200 km/s$ and mass flux $10^{-7} M_\odot/yr$ was imposed at each
timestep on the cells of the grid within an origin-centered sphere of
radius equal to 10\% the radial size of the simulation.  We chose an
environment given by \citet{HartmannCalvetBoss96} derived from
their simulations of a collapsing, rotating, axially symmetric sheet.
In this model, the collapse of the initially-flattened protostellar
cloud gives rise to a highly anisotropic density distribution in the
infalling envelope, which in turn helps strongly collimate the
initially spherical flow.  As Hartmann et al. point out, whether or
not this model is correct in detail, the general property of
non-spherical clouds to flatten or become more anisotropic as
gravitational collapse proceeds lends credence to the idea of
aspherical infall.  The anisotropic stresses of magnetic fields can
also produce non-spherical protostellar clouds, and
\citet{LiShu96} argue that such initial configurations can also
help make outflows more collimated.  \footnote{However, we note that
  the static configurations suggested by Li \& Shu may not be
  particularly relevant to outflow sources, because some collapse must
  have already occurred to produce the central mass responsible for
  the outflow.  As shown by \citet{HartmannEtAl94}, even if the
  initial density configuration is not toroidal, or
  magnetically-dominated, a toroidal density distribution will result
  from collapse.}

The density is given by the equations (8), (9) and (10) in
\citet{HartmannCalvetBoss96} which modifies calculations by
\citet{CassenMoosman81} and \citet{TSC84} (referred to
hereafter as TSC).  The velocity distribution is taken from
\citet{Ulrich76}.  The equations are rewritten in a slightly
different but equivalent form in the appendix.  These equations
produce a flattened infalling toroidal density distribution with an
equator to pole density contrast $\rho_e/\rho_p > 1000$.  This is
larger than what is produced by the \citeauthor{CassenMoosman81} and
TSC models.  The TSC model was used by \citet{WilkinStahler98}.

The environment near the center of the sheet models used in this paper
is a torus of a high equatorial density ($6\times 10^8\;m_H/cm^3$)
with a half-maximum opening angle of almost 180 degrees.  In the
hydrodynamic collimation simulations of Frank \& Mellema and Mellema
\& Frank a "fat" torus was used with opening angles of $\sim90^\circ$,
similar to those obtained by \citet{LiShu96} for
magnetized collapse.  These previous simulations
produced outflows with strong collimation.  It is noteworthy that, as
we shall see in \S~{\ref{sec:Results}}, it is possible to produce
strongly collimated flows even with the wide opening angles of the
sheet distribution.

We investigate the influence of ram pressure balance between wind and 
ambient gas by holding the wind speed constant and varying the ratio of 
the inflow mass flux to the wind mass flux (denoted $f'$:)

\begin{equation}
f'={\frac{\dot{M}_i}{\dot{M}_w}}.
\end{equation}

\noindent We chose to hold the wind velocity constant because young
stellar object wind and jet velocities are well observed to be a few hundred
kilometers per second \citep{Bachiller96,ShuRudenLadaLizano91},
while the range of the mass flux ratio f' is less well known.  We have
focused on models with $f' = 10, 20, 30, 40,$ and $50$.  This was
accomplished by adjusting the infall mass flux and fixing the outflow
mass flux.  We also simulated cases where the inflow mass flux was
reduced relative to the wind, and obtained basically the same results
with only minor differences.  We note that 
a measure of the effect of varying $f'$
can be seen by calculating the angle $\theta_e$ at which the ram
pressures are equal in our simulation.  For $f' = 10$ we find $\theta_e
= 1.568$ radians (measured from the axis) which is less than $6\%$ of
a grid cell.  For $f' = 50$ we find $\theta_e = 1.519$ radians
which is less than $130\%$ of a grid cell.  Thus is is clear that
the wind will be able to push all material away from the equator in
the $f'=10$ case while $f' = 50$ simulations should experience some
material attempting to cross the inner wind sphere.

%\placetable{table:simparam}
\begin{center}
\begin{deluxetable}{cc}
\tablecaption{Model parameters. \label{table:simparam}}
\tablewidth{0pt}
\tablehead{\colhead{parameter} & \colhead{value}}
\startdata
                                              &$4.26\times 10^{15}\;cm$ radius by\\
\raisebox{1.5ex}[0cm][0cm]{Simulation Domain}&$8.52\times 10^{15}\;cm$ along axis\\ 
%\tablevspace{5pt}
                                       &256 cells radially by\\
\raisebox{1.5ex}[0cm][0cm]{Resolution}&512 cells along axis\\ 
%\tablevspace{5pt}
Wind velocity, $v_w$&$200\;km/s$ \\ 
Wind mass flux, $\dot M_w$&$10^{-7}\;M_\odot/yr$ \\ 
Central mass, $M_*$&0.21 $M_\odot$ \\ 
%\tablevspace{5pt}
                                                         &$10^{-6}\;M_\odot/yr$ $\times\;$\\
\raisebox{1.5ex}[0cm][0cm]{Infall mass flux, $\dot M_a$}&$(f'=10,20,30,40,50)$\\
%\tablevspace{5pt}
Collapse radius, $r_0$&$5.37\times 10^{16}\; cm$ \\ 
Flattening parameter, $\eta$&$2.5$ \\
Centrifugal radius, $R_c$&$4.28\times 10^{14}\;cm\;=\;28.5\;AU$
\enddata
\end{deluxetable}
\end{center}

We note here the basis of our attack on this problem. The theoretical
issue of the interaction of a stellar wind with an infalling
environment is complex. There are a number of parameters controlling
both the wind and the infalling environment.  In choosing to address
the issues involved one must ask which parameters are important
enough, a-priori, to warrant attention? Which connect directly to
issues raised in previous studies?  Which allow for a numerically
clean solution, i.e. one whose specification of initial and boundary
conditions do not impose serious transients or artifacts on the flow?
In taking on this problem we have adopted a strategy which, hopefully,
allows to us address important aspects of the physics while leaving
others for future works.  In the present paper we focus primarily on a
relatively simple treatment of initial/ boundary conditions (e.g. the
wind and infalling environment) in order to clarify how a full
solution to the governing equations differs from the more idealized
solutions explored in previous works.  We have constructed a focused
exploration of a set of numerical experiments that address key issues
not explored in previous studies.  In future works we plan to open
other dimension of the parameter space including a more detailed
treatment of aspherical winds.

\section{Results}
\label{sec:Results}

\subsection{Spherical Winds/Aspherical Environments}
\label{sec:Results-S}
The sampling in $f'$ described in the previous section can be thought
of as a sampling from a relatively strong wind to a relatively weak
wind, as the results will show.  In addition to exploring different
steady wind cases, the simulations serve as a crude initial
exploration of what might be expected in models with fully
time-dependent wind mass loss rates.  The details of time-dependence
are particularly important for FU Orionis stars
\citep{HartmannKenyon96} where the observations show considerable
variation in the wind, likely related to large fluctuations in the
accretion rate of material through the disk.  In this section, we
start with a general overview of the results and continue by
considering each simulation in more detail.

\begin{figure}[pt]\begin{center}
  \includegraphics[width=0.9\textwidth]{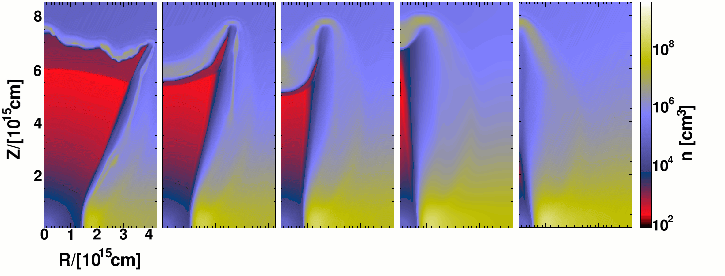}
  \caption[Strong to Weak Wind Comparison]{Comparison between
    bubbles at times of similar axial extent.  
    The density of the $f'=$10,20,30,40, and 50
    (left to right - strong wind to weak wind) cases are shown at
    their full heights at 100, 200, 220, 240, and 220 years,
    respectively.  The inner wind boundary makes a quarter circle
    around the origin with radius $4.26\times10^{14} cm$.  Note the
    change in aspect ratio of the bubbles as the wind becomes
    weaker.\label{fig:comparison}}
\end{center}\end{figure}

The principal results can be inferred from Fig. \ref{fig:comparison}
showing density snapshots of all five simulations.  In the $f' = 10$
case we see the strong wind overwhelms the infall and creates a wide
conical cavity.  As $f'$ decreases, the wind becomes weaker, the
infall mass flux dominates and, as a result, the wind becomes more
confined and collimated.  Even in the $f'=50$ case however the wind
is still strong enough to avoid being completely cut off.  In
simulations where we increase the infall flux such that $f'=100$, the
wind is completely choked by the infall with its momentum slowly
diffusing into the surrounding environment.  Because of the diffusion
and the direct interaction of the infall with the artificial wind
boundary, we are not confident in the simulations after the wind
becomes choked.  Therefore we have only placed a lower limit ($f'=50$)
on the infall to wind flux ratio at which the wind is cut off.

%\placefigure{fig:comparison}

\begin{figure}[pt]\begin{center}
  \includegraphics[width=0.9\textwidth]{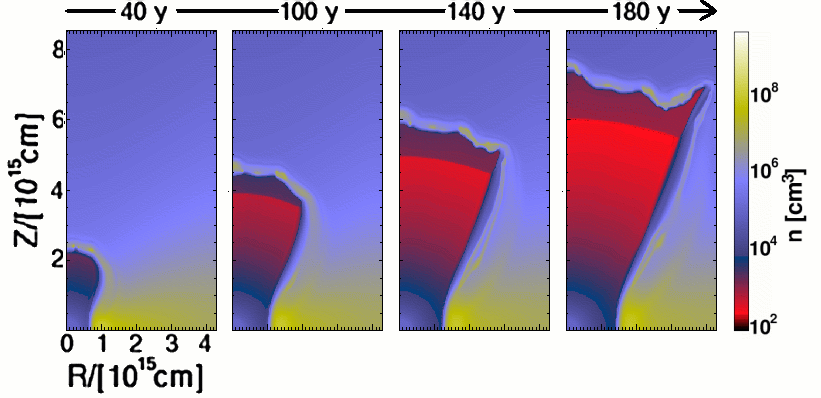}
  \caption[Strong Wind Evolution]{The density evolution of the $f'=10$ (strong wind) case.  The bubble
  begins spherical but breaks out into a cone.  Other features such as
  a partially radiative shock and conical ``jets'' appear (see
  text).\label{fig:strongEvolve}}
\end{center}\end{figure}

\begin{figure}[pt]\begin{center}
  \includegraphics[height=0.5\textheight]{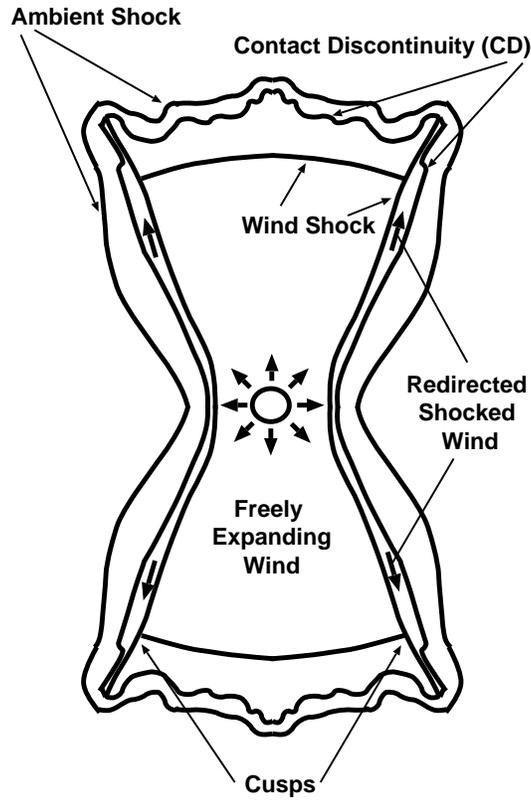}
  \caption[Diagram of Strong Wind Bubble Features]{A cartoon of the $f'=10$ (strong wind) case, labeling the
  location of the important fluid-dynamics features mentioned in the
  text.  The central source and freely expanding wind are shown.  The
  major surfaces of discontinuity (shock, contact) are identified as
  well as the sharp cusps in the wind shock.  The flow of the
  redirected wind is shown schematically.\label{fig:strongCartoon}}
\end{center}\end{figure}

\begin{figure}[pt]\begin{center}
  \includegraphics[width=0.9\textwidth]{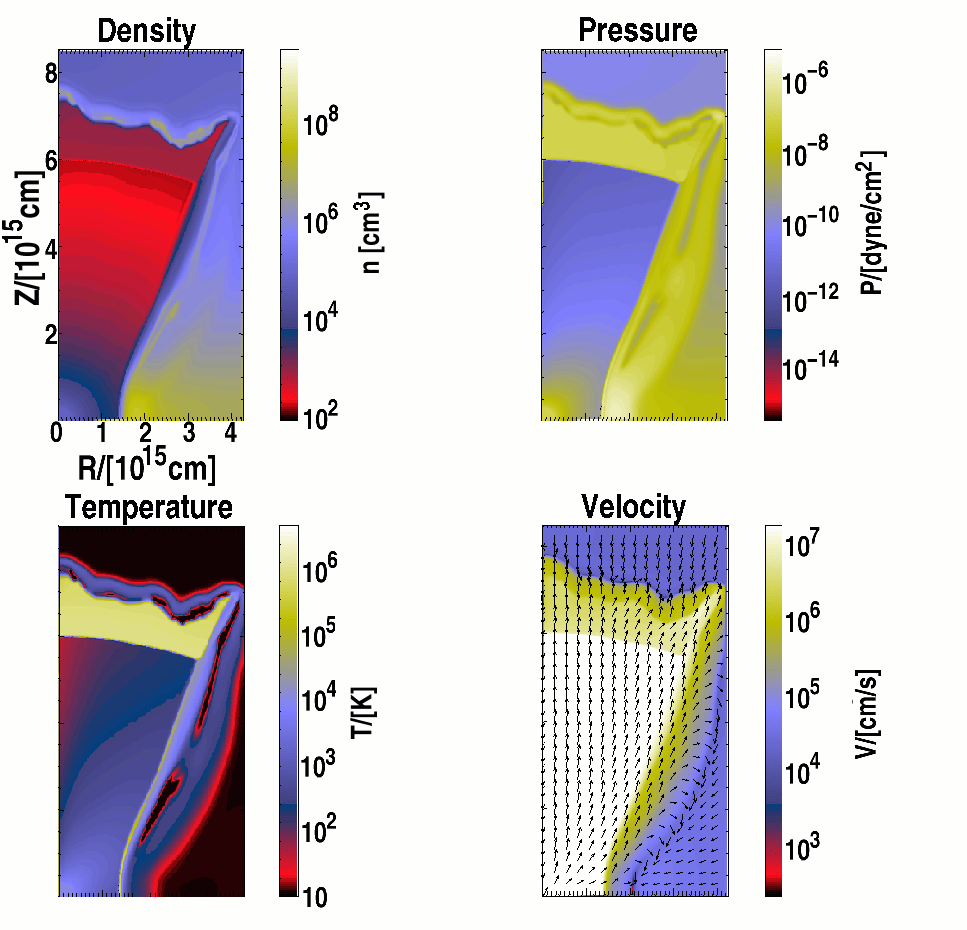}
  \caption[Strong Wind Snapshot]{The density, pressure, temperature, and velocity of the $f'=10$
  (strong wind) case at 180 years.  The velocity magnitude is
  represented by grayscale shading and the direction is represented by
  overlaid normalized arrows.  Note that in the freely expanding wind
  region (white) the velocity vectors have been sampled at regular
  grid points and fool the eye into seeing a precollimated wind.  The
  wind in that region is actually expanding spherically.
  \label{fig:strongDPTV}}
\end{center}\end{figure}

In the strong wind, ($f'=10$), case (see Fig.
\ref{fig:strongEvolve}), the wind pushes its way through the
environment but is still shaped by it.  By the time the swept up
ambient material reaches the top of the grid (at $\sim 180 ~y$: see
Figs. \ref{fig:strongCartoon} and \ref{fig:strongDPTV}) it has
a conical shape reminiscent of the outflows seen in
\citet{VelusamyLanger98} and \citet{ChandlerEtAl96}.  The
outflow bubble opens with an angle of about $30^\circ$ to the
vertical.  The supersonic wind and molecular material are each
compressed at the inner and ambient shocks respectively, forming a
thin layer about the contact discontinuity at $20^\circ$ to the
vertical.  The dynamics of this region are dominated by the fact that
the radially directed wind material hits the inner shock obliquely.
Post-shock wind is redirected and forced to flow along the contact
discontinuity.  Eventually this high speed gas flows up the walls of
the conical cavity and reaches a cusp at top of the bubble.  There the
focused high speed flow moves ahead of the bubble, encircling its nearly
spherical cap with a conical ``jet''.  While the development of such a
pattern is readily understood in terms of basic radiative oblique
shock physics it is not clear that the cusps and resulting jets would
be stable in real three dimensional bubbles.

It is interesting that as the bubble evolves, a region of warm gas ($T
\approx 10^5 ~K$) develops at the top pushing the inner and outer
shocks apart there.  Usually wind blown bubbles are classified as
either adiabatic or radiative depending on whether the cooling time is
longer or shorter than the expansion time of the bubble.  We believe
that this top region does not fall into either category, and that
instead we are seeing the development of a partially radiative bubble.
This third classification, which lies between the
adiabatic and radiative cases, was originally explored by
\citet{KooMcKee92B}.  Such a case occurs when the wind material
resupplies thermal energy generated at the shock faster than radiative
cooling can remove it, but when the cooling time is still shorter than
the bubble expansion time.  To see if the system is partially
radiative at the top of the $f'=10$ bubble we need to compare the
cooling time $t_c$ with the wind crossing time $t_x$.

Given that the top of the bubble has an essentially spherical shape we
can use the analysis of 1-D bubbles to write the cooling and crossing
times in terms of the distance from the wind source, $R$, plus other
simulation parameters for the $f'=10$ case.  A rough estimate of the
inner shock speed is $v_s = v_w=200 ~km/s$ which produces a
temperature of $8.8\times 10^5 K$. The Dalgarno-McCray cooling curve
at this temperature gives a $\Lambda(n,T)/n^2$ of $1.65\times
10^{-22}\; erg\;cm^3/s$.  Using $\dot M_w=10^{-7} M_\odot/yr$ and the
factor of 4 density increase behind a strong shock we can determine
the cooling time from the following equation:

\begin{equation}
t_c=\epsilon_{th}/\Lambda(n,T)\sim 4100 s\;(R/AU)^2.
\end{equation}

\noindent The crossing time, $t_x$, is given simply in terms
of the radial distance from the origin to the shell and the wind
velocity:

\begin{equation}
t_x={{R}\over{v_w}}\sim 7.5\times 10^5 s\; (R/AU).
\end{equation}

\noindent Note that the crossing time grows linearly and the cooling
time grows quadratically.  Thus the cooling time eventually surpasses the
crossing time.  For the parameters given above this occurs when the
shock is at a distance of about $182 ~AU$ or $2.7\times 10^{15} ~cm$
which is well inside the simulation domain.  We find that this is
approximately the distance where the simulated bubble begins to
exhibit its cap of warm postshock wind material.  We believe that
this is the first time that the partially radiative bubble phase has
been seen in a simulation, confirming the prediction of Koo \&
McKee.


The weaker $f'=20$ case is seen in Fig. \ref{fig:comparison} to also
develop a conical outer shock, but the overall shape is more confined
or collimated than in the previous case.  The cusp which forms at the
edge of the bubble is more pronounced.  This effect may due to higher
inertia of the shocked wind flow.  At low latitudes material is
channeled from regions closer to the origin where the density is
higher.  In addition, the polar regions of the bubble are expanding
slower due to the relatively lower wind mass loss rate.  Thus the
refracted wind material emerges from the top of the bubble with higher
momentum relative to the polar cap in this model than in the $f' = 10$
case, allowing the cusp to propagate farther from the cap.  Another
difference at the pole is that less warm post-shock gas exists in this
case.  This effect can also be attributed to the the slower speed of
the polar sectors of the bubble.  A slower outer shock speed implies a
faster inner shock speed (in the frame of the wind).  This in turn
leads to higher post-shock temperatures and stronger cooling in the
shocked wind.  Note that the inner shock forms an angle of about
$10^\circ$ to vertical and the outer shock (bubble) opens roughly to
$20^\circ$.

\begin{figure}[pt]\begin{center}
  \includegraphics[width=0.9\textwidth]{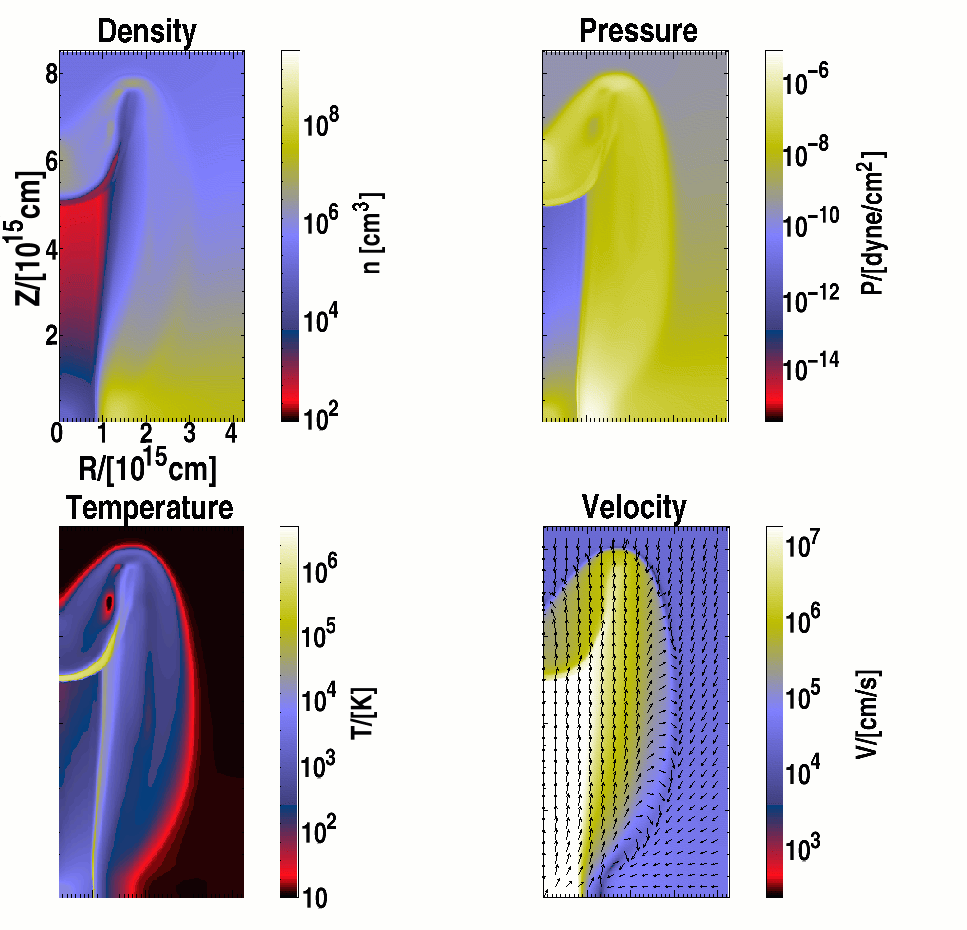}
  \caption[Moderate Wind Snapshot]{The density, pressure, temperature, and velocity of the $f'=30$
    (moderate wind) case at 220 years.\label{fig:moderateDPTV}}
\end{center}\end{figure}

At weaker winds ($f'=30$) the inner shock becomes more confined and
the flow is even more collimated.  The cusp and resulting conical
``jet'' is more pronounced than seen previously.  Again this is
expected because the cap of the bubble will expand at lower
velocities.  Figure \ref{fig:moderateDPTV} shows the details of the
flow pattern for this case.  Since the $f' = 20$ and $f' = 30$ models
are similar the figure emphasizes the points concerning the flow pattern
made above.  The wind shock opens about $2^\circ$ to vertical and the
bubble opens about $20^\circ$.

In the second weakest case ($f'=40$) the inner shock becomes almost
elliptical or ``bullet'' shaped.  The tip of the shock at the axis is
suspect because of the imposition of reflective boundary conditions
and may not form if the cylindrical symmetry is relaxed (see for
example \citealt{StoneNormanIII}).  The wind is strongly focused by the
oblique inner shock producing an almost vertically directed high
velocity flow.  The resulting bubble is very well collimated.  In this
case the conical cusp is not as pronounced and the dynamics look very
much like a jet.  At the tip of the outflow (along the axis) the the
inner shock is acting in the same manner as a jet shock, decelerating
vertically directed wind and shunting it off in the direction
transverse to the propagation of the outflow.  The full opening angle
of the outer shock of the bubble is about $35^\circ$ ($17.5^\circ$ to
vertical).

\begin{figure}[pt]\begin{center}
  \includegraphics[width=0.9\textwidth]{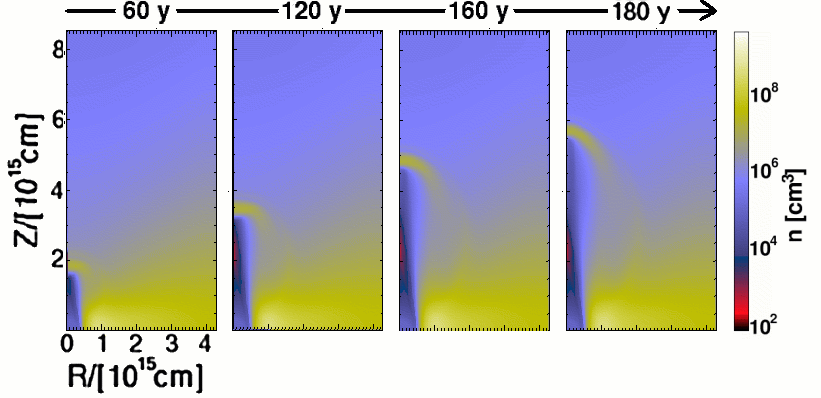}
  \caption[Weak Wind Evolution]{The density evolution of the $f'=50$ (weak wind) case.
Confinement is strong, and is maintained by a prolate
shock.\label{fig:weakevolve}}
\end{center}\end{figure}

\begin{figure}[pt]\begin{center}
  \includegraphics[width=0.9\textwidth]{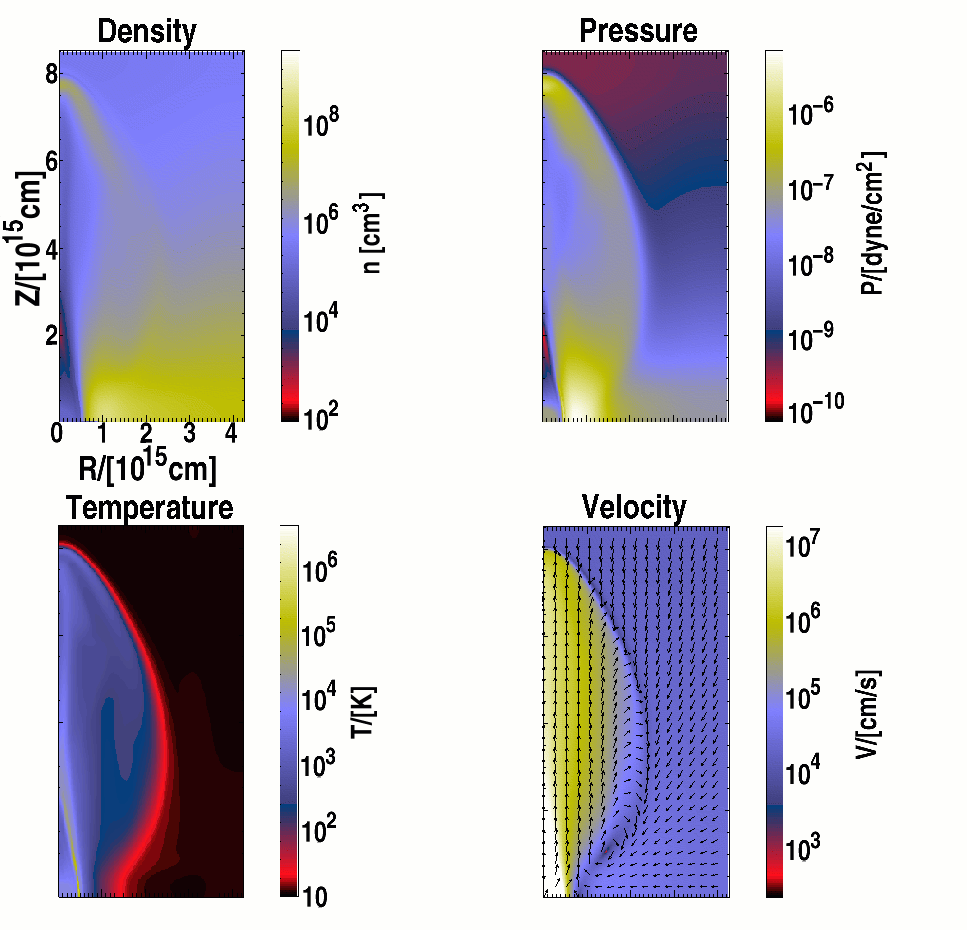}
  \caption[Weak Wind Snapshot]{The density, pressure, temperature,
  and velocity of the $f'=50$ (weak wind) case at 220 years.  Note the
  focusing effect of the inner shock evident in the velocity plot, and
  the wide region of vertically directed flow. \label{fig:weakDPTV}}
\end{center}\end{figure}

The weakest wind case ($f'=50$) is shown in Figs.
~\ref{fig:weakevolve}, ~\ref{fig:weakDPTV}.  It exhibits very strong
collimation.  The inner shock is strongly prolate and closes back on
itself at a relatively small distance from the central source, and
redirects all the wind into the collimated outflow.  There is no
extended region of freely expanding wind.  This occurs because
inflowing material is overwhelming the wind at low latitudes,
inhibiting its expansion everywhere except at the poles.  Even though
the opening angle of the bubble's outer shock at the base is
$35^\circ$, the resulting outflow will appear cylindrical because of
the focusing.


The trend of the evolutionary timescales of the different simulations
allows us to infer the dynamical consequences of collimation.  The
timescales are taken from the snapshots in Fig.~\ref{fig:comparison}
which were chosen such that the bubbles are of comparable height.  As
the wind becomes weaker ($f'=10$ to $f'=50$) the timescales go
$100~y$, $200~y$, $220~y$, $240~y$, $220~y$.  The bubbles decelerate
as the wind weakens.  The effect saturates however and then reverses
as the collimation becomes stronger.  At first glance this reversal
may be unexpected because as winds become weaker they should supply
less momentum to the bubble.  However, as the wind becomes focused
into a jet the net flux of momentum, $F = \rho v^2 A$, is injected
into the environment across a smaller angular extent $A$.  Thus the
tip of a highly collimated bubble expands at a higher velocity than
would occur for a spherical bubble.

We shifted from the "strong wind" to "weak wind" cases in these
simulations by increasing the infall mass flux while fixing the wind
velocity.  If instead we shift from $f'=10$ to the $f'=50$ by
decreasing the mass flux of the wind and keeping the infall mass flux
fixed we find that the results are qualitatively the same.  The post
inner shock region expands slightly faster due to the lower density
and reduced cooling there, but the effect is not significant.  The
results are, therefore, determined solely by the parameter $f'$.

The results of the simulations show the complicated way in which the
environment shapes the outflow.  This occurs directly at the outer
shock by providing an inertial gradient along the shock face.  It also
occurs indirectly by affecting the shape of the inner shock.  Shaping
the inner shock leads to shock focusing of wind material, creates
cusps at the tips of the bubble and, in the most extreme cases, leads
to focusing into a jet through prolate inner shocks.  In all cases,
the simulations show the ``internal'' flow of post inner shock wind
material plays an important role in the dynamics.

The results also show the way in which outflows affect the
environment.  The winds carve out evacuated cavities of potentially
large angular extent and sweep ambient material at a high rate.  The
swept up mass totals $1\times 10^{-6} M_\odot/yr$ to a few times
$10^{-7} M_\odot/yr$ as the wind becomes weaker.  Because the winds
become more collimated and on the whole expand much more slowly as
they weaken, ambient mass is swept at a slower rate.  The wider angle
winds are able to collect mass much more rapidly even though the
envelope has less material in it.

\subsection{Spherical Winds/Aspherical Environments}
Although they are not the focus of this study, we have included the
results of simulations with aspherical winds injected into a spherical
environment for comparison.  We note that addressing the issue of aspherical 
winds means specifying a model for those winds.  This
raised a number of questions suh as; should simple sinusoidal variation be used 
or something akin to the cylindrical stratification seen in wide-angle models 
like those of the X-winds?
If we focus on the former should we vary only the density from pole to equator 
or only the velocity.  If they both vary then should momentum or energy flux be 
held constant across the stellar surface?  Given these the complexities
we have chosen a highly simplified set of boundary conditions for these models
and present them as means of contrasting the results of the previous section.

The distribution in the environment is given by \citet{TSC84} with the same 
infall parameters as used in the
strong-wind simulation described above: essentially it is the
previously described sheet distribution except unflattened.  The wind
velocity distribution is given by

\begin{equation}
  \label{eq:narrowwind}
v_w
(\cos\theta)^{\log(\cos\theta_c) /log\frac{1}{2}},
\end{equation}
where $\theta$ is the polar angle, $v_w$ is the maximum polar
$200$ km/s velocity and the angle of half-maximum velocity is
$\theta_c$.  The wind mass density at every point is kept the same as
in the strong-wind simulation.

\begin{figure}[t]
\begin{center}
\begin{minipage}{0.5\textwidth}
  \includegraphics[width=0.7\textwidth]{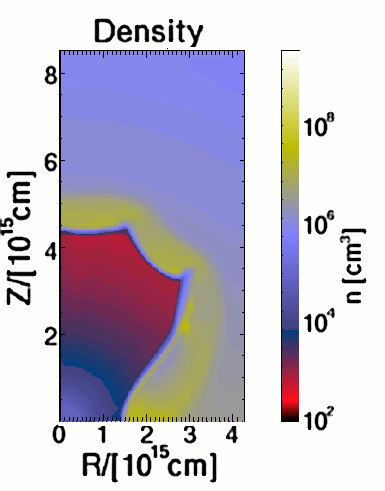}
  \caption[$70^\circ$ Snapshot]{\small The density at 300 years of a case where a non-spherical
  wind is injected into an isotropic infalling environment.  The
  half-maximum velocity is at $\theta_c=70^\circ$ from the pole.  Note
  the small extent of the bubble at this late time and the minor
  focusing. \label{fig:wide70}}
\end{minipage}
\end{center}
\end{figure}

The density of the result of injecting a mildly focused wind
($\theta_c=70^\circ$) is shown in Figure ~\ref{fig:wide70}.  Even at
$300$ years, the bubble is much more limited in size than in the sheet
environment cases: the expansion speed is lower and very little warm
gas exists to push the bubble outwards.  The focusing is also much
more limited because of the the environment.

\begin{figure}[t]
\begin{center}
\begin{minipage}{0.5\textwidth}
  \includegraphics[width=0.7\textwidth]{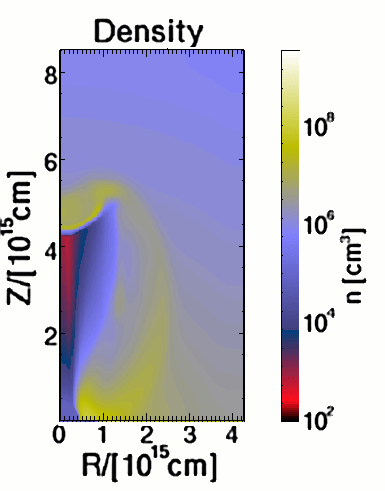}
  \caption[$30^\circ$ Wind Snapshot]{\small The density at 300 years of a case where a narrow wind
  is injected into an isotropic infalling environment.  The
  half-maximum velocity is at $\theta_c=30^\circ$ from the pole.  The
  outflow is collimated as expected.    Here, as in the wider wind
  case in Figure ~\ref{fig:wide70}, we find that the bubble is growing 
  much more slowly than those formed in the collapsing sheet
  environment.  The infalling material is reaching the wind sphere in
  this case, so the exact shape of the bubble obtained in this
  simulation is suspect. \label{fig:wide30}}
\end{minipage}
\end{center}
\end{figure}

The resulting density snapshot of injecting a more focused wind
($\theta_c=30^\circ$) into the spherical environment is shown in
Figure ~\ref{fig:wide30}.  Here the focusing is pronounced, but the
bubble extent at 300 years is still comparatively small because of
the greater mass in the polar regions.  As is to be expected, a
focused wind in a spherical environment can also produce a focused
outflow.  Detailed results, however, such as the precise bubble shape
should not be taken as final because the infall material is able to
press up against the artificial wind boundary.

These simulations show that the shape of a wind blown bubble is not
unique to a single set of boundary conditions.  As one might expect we
{\it do} see collimated outflows emerging from collimated winds though there
are some differences from the spherical wind case in terms of
propagation speed and post-shock temperatures.  These results show that
the issue is always one of wind momentum or ram pressure injection as a 
function of angle.  With spherical-winds/aspherical-infall shock focusing helps
redirect the wind ram pressure towards the pole.  This is explored in more
detail in the next section. In the aspherical winds/spherical-infall
case the wind momentum must be strongly collimated a-priori to produce
a narrow outflow.

Thus flow shapes are not unique nor should one expect them to be.
However, as we will explore in the next section, the simulations
presented in section \ref{sec:Results-S} demonstrate an effectiveness
of wide-angle wind models not seen in previous studies.

\section{Discussion}
In this section we discuss the results of our simulations in light of
previous analytical models of YSO wind blown bubbles We also attempt
to establish some points of contact between our models and molecular
outflow observations.  In making these comparisons we must note a
limitation of the present simulations, namely that they formally refer
to only very small spatial and temporal scales.  The overwhelming
majority of observed molecular outflow structures have typical
lifetimes of $10^4 - 10^5$~yr, in contrast to the $10^2$~yr time
sequences shown here.  In addition, for flows that are much larger
than $0.1$~pc, it is very likely that the bulk of the swept-up
interstellar material originally resided in regions outside the
protostellar core, let alone the inner infall regions.  Nevertheless,
there are a couple of observations of younger and smaller outflows
\citep{ChandlerEtAl96} where our simulations may be more directly
relevant.  In addition some of the physical results may be generalized
to larger-scale situations.

\label{sec:Discussion}

\subsection{Comments on the Shu, et. al. ``Snowplow'' Outflow
Model\label{sec:Snowplow}}

In \citet{ShuRudenLadaLizano91} a simple, elegant ``snowplow''
model of the formation of bipolar molecular outflows is presented.
The basic features of this model were a central wind with a mass flux
varying with polar angle and an environment with a density that varies
with polar angle and falls off in radius as $1/r^2$.  A number of
simplifying assumptions are invoked. Cooling is assumed instantaneous
so that the bubble is a purely momentum-driven thin shell.  The
interaction of the ambient material and the wind is taken to be fully
mixed and ballistic such that the dynamics at a given polar angle are
independent of those at nearby angles.  In addition, the swept up wind
mass is taken to be negligible.

In the snowplow model, changing the ratio of total infall to wind mass
flux ($f'$) only effects the bubble's expansion timescale but doesn't
change the self-similar shape.  The $1/r^2$ ambient profile leads to a
simple expression for the bubble shape and the variation with polar
angle in the extent of the bubble is purely determined by the angular
variations of wind and ambient medium.  The simulations however show a
strong variation of bubble shape with $f'$.  This discrepancy implies
that some dynamical aspects of the simulations are not captured in the
snowplow model.  

The most important physical difference between the simulations and the
snowplow model is that the snowplow model does not account for the post
inner shock redirection of wind material.  Pressure gradients in the
post-shock shell that come about because of the variation in the
ambient density, and resulting variations in shock obliquity and shock
speed have no consequence in the snowplow model, but have profound
effect in the simulations.  If we were to somehow model the non-radial
direction of the postshock wind and use that in the snowplow model we
might obtain a better agreement.  The question becomes should one use
the flow inside or outside the inner wind shock as the input wind to
the snowplow model.  While this may seem, on one level, to be a
question of where one starts the calculation, more fundamentally it is
an issue of what initial physics one includes to determine the wind
distribution.  \citet{LiShu96} make a modification of the
snowplow model in this way by incorporating the distribution of the
wind {\em after} it has been accelerated and partially focused by
magnetic launching.

If dynamically important physics in wide wind scenarios are left out
of the snowplow model, it leads to the question whether the model is
useful in predicting observations.  \citet{MassonChernin92} and
\citet{LiShu96} have used the snowplow model to calculate
distributions of mass and velocity to argue whether or not wide angle
winds can drive outflows.  Our simulations, however, show that even if
the driver is initially a wide angle wind it can be focused by a
nonlinear interaction between the wind and environment.  Our
simulations appear to blur the distinction between wide angle winds
and jets as the sole driving mechanism of the outflow.

\subsection{Mixing and momentum transfer}

A strong assumption used in the analytic models of
\citet{ShuRudenLadaLizano91} and \citet{WilkinStahler98} is
that postshock wind and postshock ambient material become fully mixed
on a dynamically short timescale.  This allows the shell to be treated
as a single fluid and greatly simplifies the calculations.  To support
the assumption, Wilkin \& Stahler consider supersonic shear flows
within the postshock regions and state that some form of instability
will generate turbulence and rapidly mix the wind and ambient fluids.
Our simulations have some diffusive mixing of fluids due to numerical
effects, but no complete mixing occurs and there is no turbulence
modeled in the code.  Since the simulations show the post-shock shear
flow influencing the shape of the bubble, it is worth discussing when
turbulent mixing of ambient and wind material will occur and, if it
does occur, how it might affect the bubble's evolution.

\subsubsection{Turbulent length scale}

The onset of turbulence in jets and bubbles is still poorly
understood, but one estimate of the momentum transfer rate between
material in a jet and a stationary ambient medium has been performed
by \citet{CantoRaga1991}.  From this, they calculate the length,
$L_t$, a laminar jet can travel before it is subsumed in a turbulent
boundary layer finding that $L_t$ scales linearly with jet Mach
number, and inversely with the jet radius $R_j$. At Mach 10 they find
$L_{t}/R_j \approx 170$.  The calculations were performed assuming
slab symmetry and are applicable to our supersonic postshock wind
regions.  \citet{RagaCabritCanto95} extend the results to the
case where both fluids are moving and find that the spreading rate is
actually reduced.

Using the result of Cant\'o and Raga we can estimate when the flows
inside the bubble become turbulent.  For this purpose we calculate the
width of the shell of post-shock wind in a slightly aspherical
isothermal wind-blown bubble.  The calculation is simplified by the
nearly spherical shape, but is still consistent with shock focusing
because even mildly aspherical bubbles will produce focused flows
(Frank \& Mellema 1996).  If density $\rho_{ws}$ in the postshock
region is constant, the mass in that region, $M_{ws}$, can be written
in terms of the volume $V_{ws}$ as:

\begin{equation}
  \label{eq:turb1}
  M_{ws}=V_{ws}\;\rho_{ws}.
\end{equation}

\noindent If the speed of the shock $v_s$ is small compared with that
of the wind, $M_{ws}$ can also be written

\begin{equation}
  \label{eq:turb2}
  M_{ws}=\dot M_{w}\;t,
\end{equation}

\noindent where $\dot M_{w}$ is the mass flux of the wind and $t$ is
the age of the bubble.  If the shock is relatively thin and the bubble
is not too aspherical then we can approximate the volume as
$V_{ws}\sim4\pi R_{s}\Delta R_{ws}$, where $R_{s}$ is the shock radius
and $\Delta R_{ws}$ is the shock width.  In terms of the shock Mach
number $\tilde{M}$ and the preshock wind density $\rho_w$ the
isothermal condition requires that
$\rho_{ws}=\rho_{w}\tilde{M}^2=\tilde{M}^2\dot M_{w}/(4\pi R_{s}^2
v_w)$ \citep{ShuVII}.  Equating (\ref{eq:turb1}) and
(\ref{eq:turb2}) and solving for the shock width gives:

\begin{equation}
  \label{eq:turb3}
  \Delta R_{ws}=\frac{t\;v_w}{\tilde{M}^2}.
\end{equation}

\noindent Once we calculate a shock speed, we can calculate the
relative width of the bubble, $\Delta R_{ws}/R_{s}$.  For simplicity
we choose the ambient environment to fall as $1/r^2$, which, as
mentioned in the previous section, implies a constant shock velocity
of $v_s=\sqrt{(\dot M_w/\dot M_a) v_w\;(a/2)}$.  Choosing typical
values of $v_w=100\; km/s$, $\dot M_w/\dot M_a=1/10$, $\tilde{M}=100$,
$a=0.2\; km/s$, we obtain:

\begin{equation}
  \label{eq:turb4}
  \frac{\Delta R_{ws}}{R_s}=\left(\frac{\dot M_w}{\dot
      M_a}\frac{a}{2}\right)^{-\half}\frac{\sqrt{v_w}}{\tilde{M}^2}
  \sim\frac{1}{100}.
\end{equation}

\noindent Note the high Mach number is a result of the fact that in an
aspherical bubble the shock creating the shear flow will be oblique
and therefore will not strongly decelerate the wind material.  In
addition the Mach number is an inverse function of the sound speed.
The strong cooling in the postshock region keeps this speed low, which
also helps to keep the Mach number high.  The simulated shear flow has
velocity on order Mach 10 (as opposed to the compression Mach number
$\tilde{M}=100$) and we can couple equation (\ref{eq:turb4}) with
the result of Cant\'o and Raga, to find that

\begin{equation}
  \label{eq:turbscale}
  \frac{\Delta R_{ws}}{R_s}\frac{L_{t}}{\Delta R}\sim 1.
\end{equation}

\noindent This suggests that the flow becomes turbulent only on a
size comparable to that of the bubble, and that the focused tangential flow
{\em will} have time to influence the dynamics before mixing occurs.

Two features this calculation doesn't take into account further reduce
the chance that turbulence will dominate the dynamics.  First, the
continuous supply of fresh wind material along the length of the shear
flow is ignored as it was in the Cant\'o \& Raga estimate.  This
effect would allow unmixed wind material to be continually resupplied
at high latitudes, possibly at a rate faster than the turbulence
incorporates material.  Second, magnetic fields are ignored.  Fields
may thread the shell especially when the bubble is small.  There is
some evidence that toroidal fields, which are present in
magnetocentrifugal wind launching models, retard the entrainment of
envelope material \citep{RosenEtAl1999}.  Poloidal fields at
moderate Mach numbers can also become wrapped up in vortices in the
mixing fluids, strengthening the field which in turn stabilizes the
shear flow \citep{FrankEtAl1996}.

We have neglected effects in our calculation that may play a mediating
role.  The most important lie in the crude way in which we treat the
geometry of the non-spherical shock.  We also haven't considered the
early transition from a spherical to an elliptical bubble, when the
shear flow Mach number will be more modest.  Also, when the shear is
stronger, centrifugal forces in the bubble may be important.  The
details of what instabilities form and how they grow in these
geometries could affect the growth rate of the turbulence.  Thus while
the estimate (\ref{eq:turbscale}) is suggestive, how quickly the flow
becomes turbulent merits further study.

\subsubsection{Direction of the turbulent flow}

Taking the opposite view, we can assume that turbulence will be
important dynamically at some time.  How then would the resulting
mixed fluid behave?  \citet{WilkinStahler98} assume the postshock
flow is dominated by the shocked ambient gas and will be carried down
towards the disk.  While there is much more mass in the ambient
material, the momentum in the wind material is quite high and the flow
could be driven poleward.  Below we calculate the direction a mixed
flow would take in a wind blown bubble shell using a simple model.

%\placefigure{fig:mixingCartoon}
\begin{figure}[pt]\begin{center}
  \includegraphics[width=0.9\textwidth]{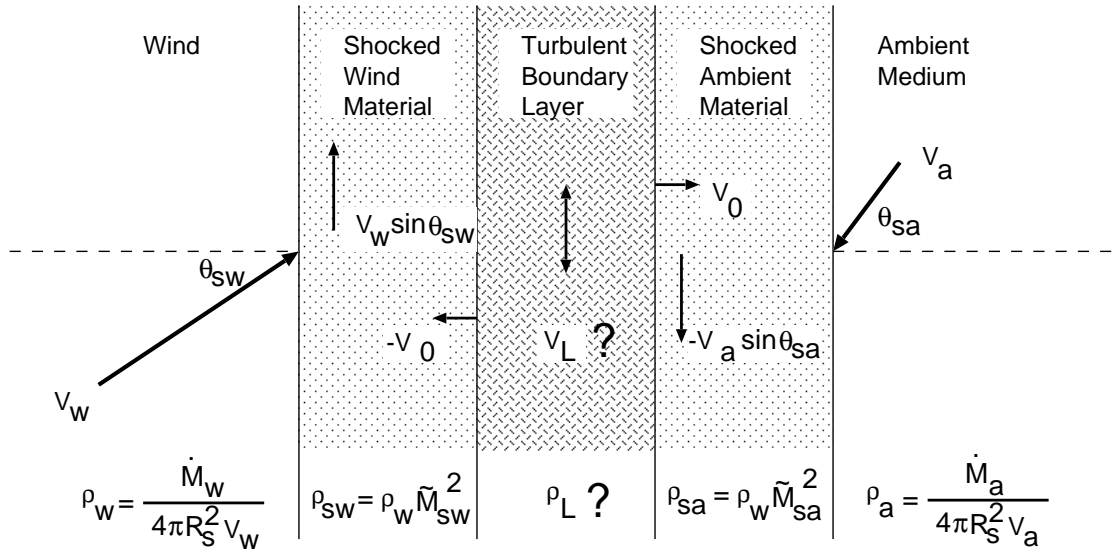}
  \caption[Mixing Model Diagram]{Cartoon for a mixing model.  The diagram represents a small
    section of the bubble's shell with wind and ambient material
    encountering shock layers at an oblique angle.  The central mixing
    layer subsumes material from postshock wind and ambient material
    which both have velocities parallel to the layer interfaces.  The
    text describes the calculation of the direction of the velocity in 
    the mixing layer.  For parameters of interest, we find that the
    wind determines the direction of the mean velocity of the
    turbulent region.\label{fig:mixingCartoon}}
\end{center}\end{figure}

In figure \ref{fig:mixingCartoon} we show preshock and postshock
regions for the ambient and wind material with a mixing layer
sandwiched between.  These regions denoted by subscripts $a$,$w$, and
$L$ respectively.  We assume the infall $v_a$ and wind velocities
$v_w$ are constant and the turbulent boundary layer grows into the
postshock regions at the same constant velocity $v_0$.  In this
framework the question becomes: is $v_L$ positive (\ie poleward) or
negative (\ie equatorward)?

Mass conservation implies:

\begin{equation}
  \label{eq:mixMassConservation}
  \rho_L 2 v_0 =\rho_{sa}v_0 -\rho_{sw}(-v_0)\;\implies\;
  \rho_L=\half(\rho_{sw}+\rho_{sa}).
\end{equation}

\noindent Likewise, momentum conservation for the component
moving parallel to the faces of the boundaries implies

\begin{equation}
  \label{eq:mixMomentumConservation}
  (\rho_L v_L) 2 v_0 =(\rho_{sa} v_{sa}) v_0 - (\rho_{sw}
  v_{sw})(-v_0)\;\implies\;
  2 \rho_L v_L=\rho_{sa} v_{sa}+\rho_{sw}v_{sw}.
\end{equation}

\noindent The preshock densities are determined from the mass flux and
velocity through $\rho=\dot M/(4\pi R_s^2 v)$, assuming the shell is
thin.  By taking the shocks to be isothermal, we obtain the postshock
densities by multiplying the preshock densities by $\tilde M^2$ where
$\tilde M$ is the Mach number.  Because the fluid velocity parallel to
the shock remains unchanged through the front, $v_{sa}=-v_a
\sin\theta_{sa}$, and $v_{sw}=v_w \sin\theta_{sw}$.  Incorporating
these velocities and solving the equations
(\ref{eq:mixMassConservation}) and (\ref{eq:mixMomentumConservation})
for $v_L$ gives:

\begin{equation}
  \label{eq:mixVelocity}
  v_L=\frac{1}{2\rho_l}\frac{\dot M_a\tilde{M}_a^{\;2}}{4\pi
    R_s^2}
  \sin\theta_{sa}
    \left(\frac{\dot M_w}{\dot M_a}
      \left(\frac{\tilde{M}_{sw}}{\tilde{M}_{sa}}\right)^2
      \frac{\sin\theta_{sw}}{\sin\theta_{sa}}-1\right).
\end{equation}

\noindent We are most interested in whether the mixed flow moves in
the direction of the wind or the infall.  If the wind and ambient
velocities are mostly aligned across the shell, then
$\sin\theta_{sa}=\sin\theta_{sw}$, and the condition for the mixed
flow moving in the direction of the wind is:

\begin{equation}
  \label{eq:mixUpCondition}
  \left(\frac{\dot M_w}{\dot M_a}
      \left(\frac{\tilde{M}_{sw}}{\tilde{M}_{sa}}\right)^2
  \right)>1.
\end{equation}

\noindent For typical values of $\tilde{M}_{sw}\sim100$,
$\tilde{M}_{sa}\sim10$, ${\dot M_w}/{\dot M_a}=1/10$, the left hand
side of (\ref{eq:mixUpCondition}) is $\sim10$.  Thus, the flow moves
in the direction of the wind.

We find that the large momentum input of the wind would tend to
accelerate the mixed material poleward.  At the pole, material
collides and sprays ahead of the bubble.  This is the Cant\'o focusing
mechanism \cite{CantoRodriguez80}, called conical converging flows,
which operates in simulations under a variety of circumstances
\citep{MellemaFrank97}.  Material would be cleared and accelerated
above the pole leading to a more elongated bubble, and possibly a very
narrow jet.

Ignoring shear flows simplifies calculations greatly.  We have argued,
however, they are important dynamically in bubbles driven into
stratified environments whether turbulence is dominant or not. Future
models will have to consider such flows in more detail.

\subsection{Properties of young outflows}

As noted in the introduction, \citet{ChandlerEtAl96} point to
three qualities that models of molecular outflows must reproduce to
match their observations of TMC-1 and TMC-1A.  They require conical
outflow lobes close to the star, evacuated outflow cavities, and
moderate $30^\circ-40^\circ$ opening angles, and they argue that
existing models do not explain these effects.

We find, however, that our simulations create outflows meeting the
requirements.  The opening angle of bubbles in the $f'=20$ to $f'=50$
cases are $35^\circ-40^\circ$.  To be sure, these angles are
determined from the density profiles of the simulations and should in
the end be determined from synthetic emission maps.  However, the
angles are not as wide as might naively be expected from a wide angle
wind.  The similarity to TMC-1, TMC-1A and other young outflows is
encouraging because these simulations are on a small spatial and time
scale (300 years, $10^{16}\;cm$) and would hopefully match up more
readily with younger outflows such as these.  

The opening angle of the bubble in the $f'=10$ case is $60^\circ$,
which is too wide to compare with TMC-1 and TMC-1A, but wider
outflows have been observed.  The outflow around IRS1 in Barnard 5 has
an opening angle of about $125^\circ$ \citep{VelusamyLanger98}.
Although these outflows may be explained by multiple non-aligned
driving winds or jets, very strong wide angle winds may be at work in
such situations.

\subsection{Mass vs. velocity relations.}

\citet{MassonChernin92} used CO line intensities of NGC 2071,
L1551, and HH 46-47 and assumptions about optical depth to obtain the
amount of mass per velocity channel in these outflows.  They measured
a power law in this mass versus velocity relationship of $\Delta
m/\Delta v=v^\gamma$ with an exponent of $\gamma=-1.8$.  With such a
steep slope little mass is accelerated to high velocities, leading
Chernin and Masson to conclude that any high velocity driver of the
molecular outflows must have a small cross-section and consequently
cannot be a wide angle wind.  Calculations of the observed mass vs.
velocity relation in other outflows have been calculated with similar
results but with interesting differences.  \citet{ChandlerEtAl96}
have pointed out that it is hard to justify fitting the distributions
in their observations with a single power law, although they do try to
fit them with two.  They find that some ``average'' power law curve
obtains a $\gamma\sim-1.8$.  \citet{BallyEtAl99} bring up the
issue of sensitivity of the transformation from line intensities to
masses to assumptions of optical depth.  The universality of a mass
vs. velocity relationship may therefore be in question, whether the
relationship is even a power law at all, and whether the mass has been
properly calculated.  We are interested however in making some
comparison of models and observations, so it is worth discussing the
mass-velocity relation calculated from the simulations.

\begin{figure}[pt]\begin{center}
  \includegraphics[width=0.9\textwidth]{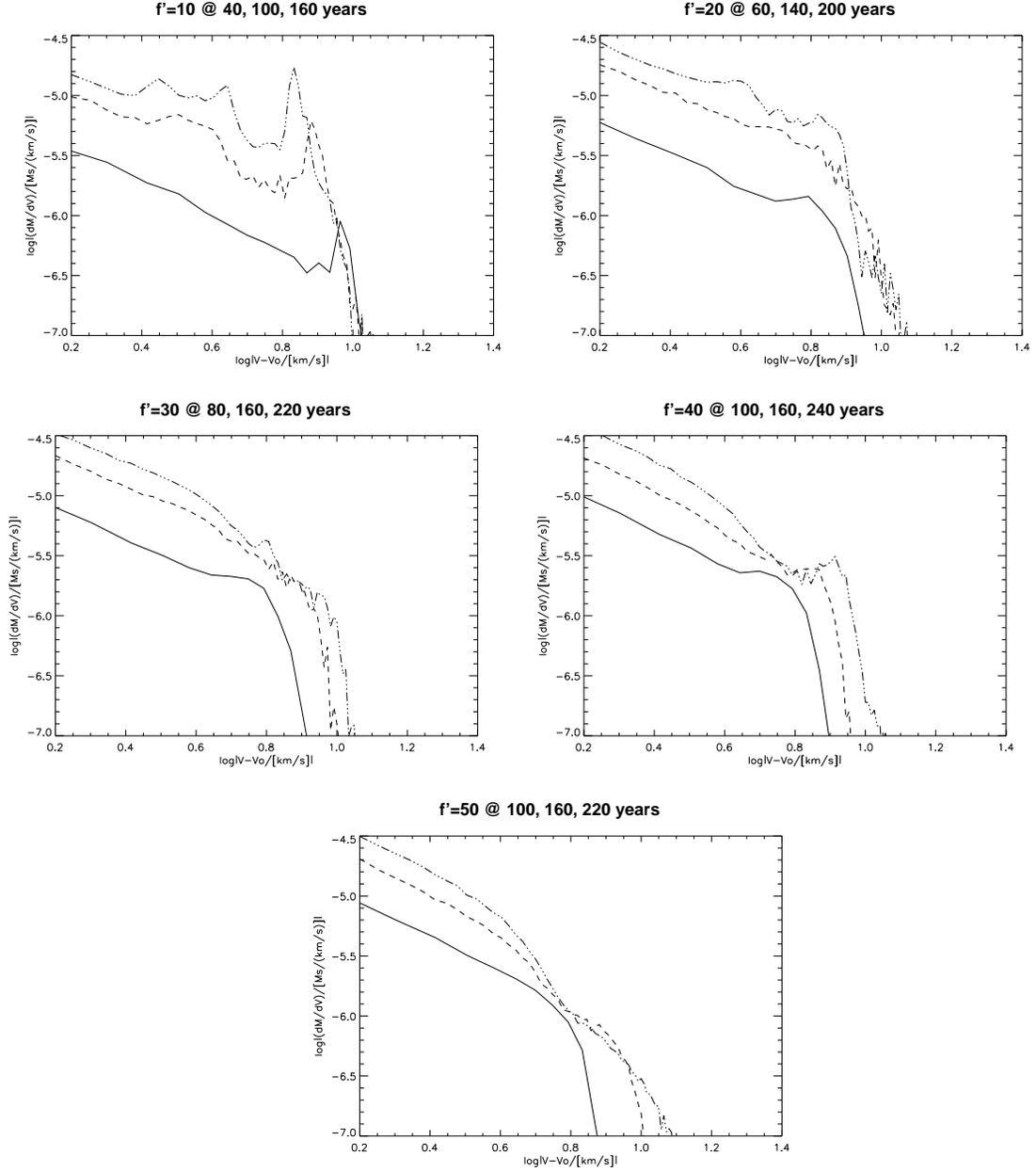}
  \caption[Mass vs. Velocity Diagrams]{Histograms of mass per line-of-sight
    (LOS) velocity channel for the simulations at different times.  As
    the simulations evolve they sweep up more mass, thus the older
    curves are above the younger curves.  The velocity channels are
    0.6 km/s wide and the LOS is $45^\circ$.\label{fig:mvsv}}
\end{center}\end{figure}

Figure \ref{fig:mvsv} shows the histograms of ambient mass as a
function of projected velocity from our simulations.  We able to
unambiguously identify the ambient material through the code's fluid
tracking and independently by the location of the contact
discontinuity.  The swept up masses are 1-2 orders of magnitude lower
than the masses shown in the TMC1 and TMC1A observations of
\citet{ChandlerEtAl96}.  However, our numerical simulations are
limited in the length of time considered and the spatial region that
can be encompassed while maintaining adequate resolution.  Thus, our
simulations correspond to sweeping up much less envelope material than
has occurred in most observed outflow regions.  Quantitatively, our
models correspond to extremely young, rare systems; qualitatively,
many of the effects we show should be present in older objects.

As found in the \citet{ChandlerEtAl96} observations the
simulation curves are not well fit by a single power law.  The curves
are logically split into two parts: a ``shallow'' portion that can be
fit on the simulation plots roughly by a power law of exponent ranging
from $-1.5$ to $-1.3$ and a steeper portion.  The shallow portion
matches well with the TMC1 and TMC1A results of -1.26 and -0.75
respectively, and is fairly close to the Chernin \& Masson value of
$\gamma\sim-1.8$. These fits are subject to some small variation from
the choice of where one portion begins and ends, and the size of the
velocity bins, but varying among reasonable choices for these only
changes the exponents by a tenth or so.  The steeper portions of the
simulation curves are fit with exponents steeper than $-10$, to be
compared with an observation of $\sim-3$ (TMC1 and TMC1A) and a
jet-model value of $-5.6$ \citep{ZhangZheng97}.  There appears,
however, to be some flattening of the steep portion as the system
evolves in the $f'=50$ case, which is the most focused model.  The
slope flattens from $-19$ to about $-7$ to $-3.25$ by $220$ years.
The focusing accelerates the polar tip and sweeps up increasingly more
mass to higher velocities.  This acceleration might also occur in
other less focused cases when the outflow ``breaks out'' of the
protostar's cloud core into much less dense material.  If this
breakout flattens the mass vs. velocity curve at high velocity then it
could explain the observationally derived curves from older outflows.

The shallow and steep portions of the mass-velocity curves are
connected by a ``hump,'' which does not appear in the observationally
determined curves.  This feature is also seen in the mass-velocity
plot of \citet{ZhangZheng97} and we find that in our simulations
it is largely due to the innermost portions of the ambient shock -
those farthest from an observer.  Optical depth effects could play a
role in translating this part of the curve to an accurate integrated
CO line profile.

The mass vs. velocity relations for the simulations look promising,
with one caveat being that questions remain about this method of
matching observations to models.  Another caveat is that extrapolating
these results to those of large-scale outflows (larger than 0.1 pc) is
not straightforward.  Large outflows are likely to involve
significantly more material from regions exterior to the cloud than
material in the infall in the infall region, and thus the simulations
here may refer only to structures on 100 - 1000 AU scales.  The
Chandler et. al. observations are more relevant in this case.  Noting
these provisos, the models have mass vs.  velocity relations whose
shallow portions are about as steep as those quoted by Chernin \&
Masson, and the steeper portions may flatten as the flow focuses or
breaks out of the cloud core.  The simulated relations have striking
similarities to the observationally derived curves despite the fact
that the model systems are driven by wide angle winds.

\section{Conclusion}
\label{sec:Conclusion}

We have performed radiative simulations of the interaction of a central
spherically symmetric wind with an infalling environment under
conditions similar to those found around young stellar objects.  We
include gravity from the protostar and rotation about the axis.  We
have examined the effect of varying the relative infall to wind mass
flux on the morphology and kinematics of the resulting outflow.

In our simulations, shaping by the toroidal YSO environment of the
wind shock, in addition to the outer molecular shock, focuses the
outflow, leading to morphologies and kinematics not explored in
earlier analytical models.  Most importantly, the outflows become
progressively more collimated as the central wind becomes weaker
relative to the infall.  Stronger winds can reverse the inflow of
material creating the wide evacuated conical cavities seen in some
observations of molecular outflows.  Weaker winds are focused more
tightly and form elongated, bipolar structures.  The simplified
snowplow model for these outflows does not explain these variations.
Thus, conclusions made from the snowplow model that wide angle winds
cannot drive molecular outflows should be reexamined.

We suggest that the effects of turbulent mixing of wind and ambient
material, which are not treated explicitly in the simulations, will
not significantly alter the results.  The mixing is shown to be a
relatively inefficient process.  The shear speed of the post-shock
wind in the simulations is high and we estimate that the poleward
focused flow will traverse a large part of the wind blown bubble's
circumference before mixing becomes significant.  If the mixing does
occur, the momentum in the wind is large enough to force the flow
towards the poles and possibly form a conical converging flow.

We find a favorable comparison of the mass vs. velocity relations
derived from the simulations to those from observations.  The shallow
portion of the curve at low velocity can be roughly fit by a power law
of exponent from about $-1.5$ to $-1.3$.  The steeper portion at high
velocity is fit by an exponent of $-10$ or steeper, although the most
focused of our simulations shows a gradual flattening.  This leads us
to speculate that when the other simulations run to breakout of the
cloud core there might be further focusing and a similar transition in
the mass vs. velocity curve.

These simulations show that directed outflows can result from wide
angle winds.  The nonlinear interaction of wind and environment leads
to dynamics that are not easily modeled outside of simulation, but
which lead to behavior similar to what is seen in young outflows.
Properties that have been previously been proscribed to jet driven
systems can be found in the wide angle winds.  Further synthetic
observations need to be performed on simulations like those in this
paper to indicate the other ways in which the wide angle wind model
offers an explanation for molecular outflows.

For low mass YSOs it is probable that wide angle winds will be
generated by some form of MHD mechanism associated with an accretion
disk \citep{OuyedPudritz97} or disk-star boundary layer
\citep{ShuEtAl94}.  Although our simulations are hydrodynamic
and involve a maximally uncollimated (\ie isotropic) wind, the
shaping of outflows should be general and occur even in the presence
of magnetic fields.  This is because shock focusing could occur even
when the shocks are magnetized.  In addition, MHD driven wide angle
wind models will likely enhance the outflow collimation in two ways.
The first is because these models typically collimate the wind early,
in the ``wind sphere'' region we do not model, to send more momentum
along the poles. The second is because toroidal fields are generated
and carried out by the wind in such models, providing radially
directed hoop stresses in the wind post-shock region.

In future papers we hope to explore in more detail the emissive
properties of wide wind driven simulated outflows and make a more
direct comparison with observations, as well as study long term
behavior.  Additionally, these studies of the relative magnitude of 
the wind have laid the foundation
for an examination into how time variation in the wind affects the
outflow.  With time dependence of the source, which we expect on
observational grounds, the outflows can be extremely complicated
kinematically.  Finally a more detailed exploration of the parameters
associated with the asphericity of the wind should be explored
allowing a better link between its shape and the resulting properties
of the outflows.

%-------------------------------------------------------------------------
\acknowledgements

We wish to thank Thomas Gardiner for his helpful discussions. This
work was supported by NSF Grant AST-0978765 and the University of
Rochester's Laboratory for Laser Energetics through a Frank J. Horton
Fellowship and their computing facilities.

%-------------------------------------------------------------------------

\appendix
\section{Collapsing Sheet Model}
In this appendix we briefly recapitulate the equations of the sheet
density distribution described in \citet{HartmannCalvetBoss96}.

The collapsing sheet is described as a
function of spherical radius from the origin $r$ and cosine of the
polar angle $\mu$.  In spherical coordinates the mass density and
velocity is:

\begin{eqnarray}
  \label{eq:Sheet}
  \frac{\rho}{\rho_n}(\frac{r}{R_c},\mu)&=&
           \left({\rm
               sech}(\eta\mu_0)\right)^2\left(\frac{\eta}{\tanh(\eta)}\right)
                \frac{\rho_{CMU}}{\rho_n}(\frac{r}{R_c},\mu) \\
  \frac{\rho_{CMU}}{\rho_n}(\frac{r}{R_c},\mu)&=&\left(\frac{r}{R_c}\right)^{-3/2}\left(1+\frac{\mu}{\mu_0}\right)^{-1/2}
           \left(\frac{\mu}{\mu_0}+\frac{2\mu_0}{\frac{r}{R_c}}\right)^{-1} \\
  \frac{v_r}{v_k}(\frac{r}{R_c},\mu)&=&-(\frac{r}{R_c})^{-1/2}\left(1+\frac{\mu}{\mu_0}\right)^{1/2} \\
  \frac{v_\theta}{v_k}(\frac{r}{R_c},\mu)&=&(\frac{r}{R_c})^{-1/2}(\mu_0-\mu)\left(\frac{\mu_0+\mu}{\mu_0(1-\mu^2)}\right)^{-1/2} \\
  \frac{v_\phi}{v_k}(\frac{r}{R_c},\mu)&=&(\frac{r}{R_c})^{-1/2}\left(1-\frac{\mu}{\mu_0}\right)^{1/2}\left(\frac{1-\mu_0^2}{1-\mu^2}\right)^{1/2}.
\end{eqnarray}

\noindent where $\mu_0=\mu_0(r/R_c,\mu)$ specifies the location
(cosine of polar angle) on a reference sphere at some radius $r_0$
which a gas parcel at the given coordinates came from.  It is defined
implicitly by:

\begin{equation}
  \label{eq:sourceAngle}
  \frac{r}{R_c}=\frac{1-\mu_0^2}{1-\frac{\mu}{\mu_0}}.
\end{equation}

\noindent The parameter $\eta$ is the ratio of the distance inside-out
collapse has progressed into the sheet $r_0$, to the scale height $H$
of the initial, static, self-gravitating sheet.  It also describes the
degree of flattening in the central density distribution.  The value
of $\eta$ is taken as constant during the simulations.  Other basic
constants include the centrifugal radius, $R_c$, the Keplerian
velocity at the centrifugal radius, $v_k$, and a density $\rho_n$.
These in turn can be defined using the mass of the (forming) central
star $M_*$, and an angular velocity at the radius $r_0$ of
$\dot\phi_0$.

\begin{eqnarray}
  \label{eq:sheetConstants}
  R_c&=&\frac{\dot\phi_0^2\,r_0^4}{GM_*}\\
  v_k&=&\sqrt{\frac{GM_*}{R_c}}\\
  \rho_n&=&\frac{\dot M_a}{4\pi(GM_*R_c^{\;3})^{1/2}}
\end{eqnarray}

\newpage


\newpage

\begin{thebibliography}{}

\bibitem[\protect\citeauthoryear{Bachiller}{Bachiller}{1996}]{Bachiller96}
Bachiller, R. 1996,
\newblock Bipolar molecular outflows from young stars and protostars
\newblock In {\em ARAA}, Volume~34, pp.\  111-154. Annual Reviews.

\bibitem[\protect\citeauthoryear{{Bally}, {Devine}, {Alten}, \&
  {Sutherland}}{{Bally} et~al.}{1997}]{BallyEtAl1997}
{Bally}, J., {Devine}, D., {Alten}, V., \& {Sutherland}, R.~S. 1997,
\newblock \apj~{478}, 603

\bibitem[\protect\citeauthoryear{Bally, Reipurth, Lada, \& Billawala}{Bally
  et~al.}{1999}]{BallyEtAl99}
Bally, J., Reipurth, B., Lada, C.~J., \& Billawala, Y. 1999,
\newblock \apj~{117}, 410

\bibitem[\protect\citeauthoryear{{Blandford} \& {Payne}}{{Blandford} \&
  {Payne}}{1982}]{BlandfordPayne82}
{Blandford}, R.~D. \& {Payne}, D.~G. 1982,
\newblock \mnras~{199}, 883

\bibitem[\protect\citeauthoryear{{Calvet} \& {Gullbring}}{{Calvet} \&
  {Gullbring}}{1998}]{CalvetGullbring1998}
{Calvet}, N. \& {Gullbring}, E. 1998,
\newblock \apj~{509}, 802

\bibitem[\protect\citeauthoryear{{Cant\'o} \& {Raga}}{{Cant\'o} \&
  {Raga}}{1991}]{CantoRaga1991}
{Cant\'o}, J. \& {Raga}, A.~C. 1991,
\newblock \apj~{372}, 646

\bibitem[\protect\citeauthoryear{Cant\'o \& Rodr\'iguez}{Cant\'o \&
  Rodr\'iguez}{1980}]{CantoRodriguez80}
Cant\'o, J. \& Rodr\'iguez, L.~F. 1980,
\newblock \apj~{239}, 982

\bibitem[\protect\citeauthoryear{{Cassen} \& {Moosman}}{{Cassen} \&
  {Moosman}}{1981}]{CassenMoosman81}
{Cassen}, P. \& {Moosman}, A. 1981,
\newblock Icarus~{48}, 353

\bibitem[\protect\citeauthoryear{Chandler, Terebey, Barsony, Moore, \&
  Gautier}{Chandler et~al.}{1996}]{ChandlerEtAl96}
Chandler, C.~J., Terebey, S., Barsony, M., Moore, T. J.~T., \& Gautier, T.~N.
  1996,
\newblock \apj~{471}, 308

\bibitem[\protect\citeauthoryear{Chernin \& Masson}{Chernin \&
  Masson}{1995}]{CherninMasson95}
Chernin, L.~M. \& Masson, C.~R. 1995,
\newblock ApJ~{455}, 182

\bibitem[\protect\citeauthoryear{Cliffe, Frank, Livio, \& Jones}{Cliffe
  et~al.}{1995}]{CliffeFrankEtAl95}
Cliffe, J.~A., Frank, A., Livio, M., \& Jones, T.~W. 1995,
\newblock ApJ~{447}, L49

\bibitem[\protect\citeauthoryear{Dalgarno \& McCray}{Dalgarno \&
  McCray}{1972}]{DM72}
Dalgarno, A. \& McCray, R.~A. 1972,
\newblock Heating and ionization of hi regions
\newblock In L.~Goldberg, D.~Layzer, \& J.~G. Phillips (Eds.), {\em AARA},
  Volume~10, pp.\  375 Annual Reviews Inc.

\bibitem[\protect\citeauthoryear{{Frank}, {Jones}, {Ryu}, \& {Gaalaas}}{{Frank}
  et~al.}{1996}]{FrankEtAl1996}
{Frank}, A., {Jones}, T.~W., {Ryu}, D., \& {Gaalaas}, J.~B. 1996,
\newblock \apj~{460}, 777

\bibitem[\protect\citeauthoryear{Frank \& Mellema}{Frank \&
  Mellema}{1996}]{FrankMellema96}
Frank, A. \& Mellema, G. 1996,
\newblock ApJ~{472}, 684

\bibitem[\protect\citeauthoryear{Harten}{Harten}{1983}]{Harten83}
Harten, A. 1983,
\newblock JCompPhys~{49}, 357

\bibitem[\protect\citeauthoryear{{Hartmann}, {Boss}, {Calvet}, \&
  {Whitney}}{{Hartmann} et~al.}{1994}]{HartmannEtAl94}
{Hartmann}, L., {Boss}, A., {Calvet}, N., \& {Whitney}, B. 1994,
\newblock \apjl~{430}, L49

\bibitem[\protect\citeauthoryear{Hartmann, Calvet, \& Boss}{Hartmann
  et~al.}{1996}]{HartmannCalvetBoss96}
Hartmann, L., Calvet, N., \& Boss, A. 1996,
\newblock ApJ~{464}, 387

\bibitem[\protect\citeauthoryear{{Hartmann} \& {Kenyon}}{{Hartmann} \&
  {Kenyon}}{1996}]{HartmannKenyon96}
{Hartmann}, L. \& {Kenyon}, S.~J. 1996,
\newblock \araa~{34}, 207

\bibitem[\protect\citeauthoryear{{K{\"{o}}nigl}}{{K{\"{o}}nigl}}{1989}]{Konigl%
89}
{K{\"{o}}nigl}, A. 1989,
\newblock \apj~{342}, 208

\bibitem[\protect\citeauthoryear{Koo \& McKee}{Koo \&
  McKee}{1992}]{KooMcKee92B}
Koo, B.-C. \& McKee, C.~F. 1992,
\newblock ApJ~{388}, 103

\bibitem[\protect\citeauthoryear{Lepp \& Shull}{Lepp \&
  Shull}{1983}]{LeppShull83}
Lepp, S. \& Shull, J.~M. 1983,
\newblock \apj~{270}, 578

\bibitem[\protect\citeauthoryear{Li \& Shu}{Li \& Shu}{1996}]{LiShu96}
Li, Z.-Y. \& Shu, F.~H. 1996,
\newblock ApJ~{472}, 211

\bibitem[\protect\citeauthoryear{Masson \& Chernin}{Masson \&
  Chernin}{1992}]{MassonChernin92}
Masson, C.~R. \& Chernin, L.~M. 1992,
\newblock ApJ~{387}, L47

\bibitem[\protect\citeauthoryear{Masson \& Chernin}{Masson \&
  Chernin}{1993}]{MassonChernin93}
Masson, C.~R. \& Chernin, L.~M. 1993,
\newblock ApJ~{414}, 230

\bibitem[\protect\citeauthoryear{Mellema \& Frank}{Mellema \&
  Frank}{1997}]{MellemaFrank97}
Mellema, G. \& Frank, A. 1997,
\newblock MNRAS~{292}, 795

\bibitem[\protect\citeauthoryear{{Najita} \& {Shu}}{{Najita} \&
  {Shu}}{1994}]{NajitaShu94}
{Najita}, J.~R. \& {Shu}, F.~H. 1994,
\newblock \apj~{429}, 808

\bibitem[\protect\citeauthoryear{Ouyed \& Pudritz}{Ouyed \&
  Pudritz}{1997}]{OuyedPudritz97}
Ouyed, R. \& Pudritz, R.~E. 1997,
\newblock ApJ~{482}, 712

\bibitem[\protect\citeauthoryear{{Pudritz} \& {Norman}}{{Pudritz} \&
  {Norman}}{1983}]{PudritzNorman83}
{Pudritz}, R.~E. \& {Norman}, C.~A. 1983,
\newblock \apj~{274}, 677

\bibitem[\protect\citeauthoryear{{Raga} \& {Cabrit}}{{Raga} \&
  {Cabrit}}{1993}]{RagaCabrit93}
{Raga}, A. \& {Cabrit}, S. 1993,
\newblock \aap~{278}, 267

\bibitem[\protect\citeauthoryear{{Raga}, {Cabrit}, \& {Cant\'o}}{{Raga}
  et~al.}{1995}]{RagaCabritCanto95}
{Raga}, A.~C., {Cabrit}, S., \& {Cant\'o}, J. 1995,
\newblock \mnras~{273}, 422

\bibitem[\protect\citeauthoryear{{Reipurth} \& {Heathcote}}{{Reipurth} \&
  {Heathcote}}{1997}]{ReipurthHeathcote1997}
{Reipurth}, B. \& {Heathcote}, S. 1997,
\newblock IAU Symposia~{182}, 3

\bibitem[\protect\citeauthoryear{Rosen, Hardee, Clarke, \& Johnson}{Rosen
  et~al.}{1999}]{RosenEtAl1999}
Rosen, A., Hardee, P., Clarke, D., \& Johnson, A. 1999,
\newblock \apj~{510}, 136

\bibitem[\protect\citeauthoryear{Ryu, Brown, Ostriker, \& Loeb}{Ryu
  et~al.}{1995}]{RyuEtAl95}
Ryu, D., Brown, G.~L., Ostriker, J.~P., \& Loeb, A. 1995,
\newblock ApJ~{452}, 364-378

\bibitem[\protect\citeauthoryear{{Shang}, {Shu}, \& {Glassgold}}{{Shang}
  et~al.}{1998}]{ShangEtAl1998}
{Shang}, H., {Shu}, F.~H., \& {Glassgold}, A.~E. 1998,
\newblock \apjl~{493}, L91

\bibitem[\protect\citeauthoryear{{Shu}, {Najita}, {Ostriker}, {Wilkin},
  {Ruden}, \& {Lizano}}{{Shu} et~al.}{1994}]{ShuEtAl94}
{Shu}, F., {Najita}, J., {Ostriker}, E., {Wilkin}, F., {Ruden}, S., \&
  {Lizano}, S. 1994,
\newblock \apj~{429}, 781

\bibitem[\protect\citeauthoryear{Shu}{Shu}{1992}]{ShuVII}
Shu, F.~H. 1992,
\newblock {\em Gas Dynamics}, Volume~2 of {\em The Physics of Astrophysics}.
\newblock University Science Books.

\bibitem[\protect\citeauthoryear{Shu, Ruden, Lada, \& Lizano}{Shu
  et~al.}{1991}]{ShuRudenLadaLizano91}
Shu, F.~H., Ruden, S.~P., Lada, C.~J., \& Lizano, S. 1991,
\newblock ApJ~{370}, L31

\bibitem[\protect\citeauthoryear{{Stahler}}{{Stahler}}{1994}]{Stahler94}
{Stahler}, S.~W. 1994,
\newblock \apj~{422}, 616

\bibitem[\protect\citeauthoryear{{Stapelfeldt}, {Burrows}, {Krist}, \& {Wfpc2
  Science Team}}{{Stapelfeldt} et~al.}{1997}]{StapelfeldtEtAl1997}
{Stapelfeldt}, K., {Burrows}, C.~J., {Krist}, J.~E., \& {Wfpc2 Science Team}
  1997,
\newblock IAU Symposia~{182}, 355

\bibitem[\protect\citeauthoryear{{Stone} \& {Norman}}{{Stone} \&
    {Norman}}{1994}]{StoneNormanIII} {Stone}, J.~M. \& {Norman}, M.~L.
  1994, \newblock \apj~{420}, 237

\bibitem[\protect\citeauthoryear{{Suttner}, {Smith}, {Yorke}, \&
  {Zinnecker}}{{Suttner} et~al.}{1997}]{SuttnerSmithEtAl1997}
{Suttner}, G., {Smith}, M.~D., {Yorke}, H.~W., \& {Zinnecker}, H. 1997,
\newblock \aap~{318}, 595

\bibitem[\protect\citeauthoryear{{Terebey}, {Shu}, \& {Cassen}}{{Terebey}
  et~al.}{1984}]{TSC84}
{Terebey}, S., {Shu}, F.~H., \& {Cassen}, P. 1984,
\newblock \apj~{286}, 529

\bibitem[\protect\citeauthoryear{Ulrich}{Ulrich}{1976}]{Ulrich76}
Ulrich, R.~K. 1976,
\newblock ApJ~{230}, 377

\bibitem[\protect\citeauthoryear{Velusamy \& Langer}{Velusamy \&
  Langer}{1998}]{VelusamyLanger98}
Velusamy, T. \& Langer, W.~D. 1998,
\newblock Nat~{392}, 685

\bibitem[\protect\citeauthoryear{Wilkin \& Stahler}{Wilkin \&
  Stahler}{1998}]{WilkinStahler98}
Wilkin, F.~P. \& Stahler, S.~W. 1998,
\newblock ApJ~{502}, 661

\bibitem[\protect\citeauthoryear{{Zhang} \& {Zheng}}{{Zhang} \&
  {Zheng}}{1997}]{ZhangZheng97}
{Zhang}, Q. \& {Zheng}, X. 1997,
\newblock \apj~{474}, 719

\end{thebibliography}
\end{document}